\documentclass[11pt]{article}
\usepackage{amssymb,epsf,graphics,graphicx}
\newcommand\blank[1]{}

\newcommand\toline[1]{--#1}
\newcommand{\fract}[2]{{\textstyle\frac{#1}{#2}}}

\newcommand{\ri}{\right}

\newcommand{\lf}{\left}

\newcommand{\CS}{{\cal S}}

\newcommand\ZZ{{\mathbb Z}}

\newcommand\eq{\begin{equation}}
\newcommand\en{\end{equation}}
\newcommand\bea{\begin{eqnarray}}
\newcommand\eea{\end{eqnarray}}
\newcommand\nn{\nonumber}
\newcommand\ba{\(\begin{array}}
\newcommand\ea{\end{array}\)}

\newcommand\PT{{\cal P}{\cal T}}
\newcommand{\RR}{{\hbox{$\rm\textstyle I\kern-0.2em R$}}}
\newcommand{\CH}{{\cal H}}
\newcommand{\resection}[1]{\setcounter{equation}{0}\section{#1}}

\newcommand{\One}{{\hbox{{\rm 1{\hbox to 1.5pt{\hss\rm1}}}}}}

\begin{document}
\begin{titlepage}
\vskip 8cm
\begin{center}
{\Large{\bf Aspects of the ODE/IM correspondence}}
\end{center}
\vskip 2cm
\centerline{\large{Patrick Dorey$^1$, Clare Dunning$^2$,  and Roberto
Tateo$^3$}}
\vskip 0.9cm
\centerline{\sl\small $^1$Department of Mathematical Sciences, University of
Durham,}
\centerline{\sl\small Durham DH1 3L, UK}
\vskip 0.4cm
\centerline{\sl\small $^2$
Centre of Mathematical Physics, School of Physical Sciences, The University
of Queensland,}
\centerline{\sl\small Brisbane 4072, Australia}
\vskip 0.4cm
\centerline{\sl\small $^3$Dipartimento  di Fisica Teorica e sezione INFN,
Universit\`a di Torino,}
\centerline{\sl\small Via Pietro Giuria 1, 10125 Torino, Italy}
\vskip 0.5cm
\centerline{E-mails:}
\centerline{ p.e.dorey@dur.ac.uk, tcd@maths.uq.edu.au,  tateo@to.infn.it}

\vskip 1.25cm
\begin{abstract}
\noindent
We review a  surprising correspondence  between certain two-dimensional integrable models   and the
spectral theory of ordinary differential equations. Particular emphasis is given to the relevance of this
correspondence to certain problems in  $\PT$-symmetric quantum mechanics.

{\it Contribution to the Proceedings ``Recent Trends in Exponential Asymptotics",
 June 28 - July 2 (2004),
RIMS, Kyoto.}
\end{abstract}
\end{titlepage}
\setcounter{footnote}{0}
\def\thefootnote{\fnsymbol{footnote}}
%
\resection{Prelude}

This short review  is about a surprising link between integrable models
and ordinary differential
equations~\cite{Dorey:1998pt,Bazhanov:1998wj,Suzuki:2000fc,Dorey:2000kq},
its relevance in the
study of the novel  kind of non-Hermitian quantum
mechanics studied  by Carl Bender and many
others~\cite{CGM1980,BG1993,BB,Bender:1998gh,Znojil:1999qt,%
Dorey:1999uk,Dorey:2001uw,Dorey:2001hi, Bernard:2001wh,Yan:2001gp, Shin:2002vu, Sinha:2002ga,
Mostafazadeh:2002hb, Znojil:2002yr,
 Bagchi:2002yj, Bender:2002vv, Kleefeld:2004qs,BBS} and the application of
techniques from the theory of integrable models to answer a long-standing question concerning the reality
of the energy levels of a particular set of non-Hermitian operators.

In the theory of ordinary differential equations the adjective
``exactly solvable'' usually  indicates
that the problem is fully soluble, with  solutions  expressible
in terms of some previously-defined
mathematical functions. In the context of
spectral theory of second-order differential equations the
classical example is a Schr{\"o}dinger problem  with harmonic  potential:
\eq
- \psi''(x) + x^2 \psi(x) =E_n \psi(x)~~,~~~~~~~ \psi(x) \in L^2(\RR)~.
\en
 This equation  is mathematically soluble in the sense that
the eigenfunctions are given in terms of elementary functions. On the other hand, a mathematician would be
reluctant to call a Schr{\"o}dinger  equation with  potential $x^\alpha$ with $\alpha\in
\RR^+$
  exactly solvable.  However we shall shortly see
that this family of quantum  systems  is somehow
equivalent to a  two dimensional exactly solvable model: the
six-vertex model in its scaling limit. There is clearly a
mismatch between the definitions of exact
solvability in the two fields. Naively speaking, in
two dimensions there is a special  class of
statistical models for which at least one physically-interesting
quantity, the density of  free-energy in
the thermodynamic limit, can be computed exactly.
For this reason  these class of systems are given
the name of ``exactly solvable'' or
``integrable'' models. (The existence of a complete set of
independent, commuting conservation laws  is a more precise
characterisation of an integrable system, but a full
discussion of this would go well
beyond the scope of this review.)

Notice that this does not in general mean that other
quantities, such as, for example, the finite lattice
free energy or the correlation functions,
can be exactly determined.
This sounds like a negative note (and it is!!), but during more
than 30 years of integrability the
scientific community  has partially overcome this negative aspect
by introducing a number of powerful tools which allow an
integrable  system to be studied  in great detail --
works particularly relevant to the current story include
\cite{baxterbook,KBP,Bazhanov:1996dr}.
This review  is about the use of
just one these tools in the ODE context. Of course
many techniques have also been developed in the ODE framework (see for
example \cite{Sha,Bender:dr,Voros}), and
integrable models
can, and to some extent already
do~\cite{Bazhanov:1998wj,Lukyanov:2003nj}, profit from
these as well.

\resection{The IM side: the six-vertex model}
Let us start from the integrable model side, and consider an
$N\times M$ lattice model with periodic
boundary conditions and, for an  irrelevant technical
reason, $N/2$ even. On each link of the lattice we
place a spin taking one of two values: this is conveniently
denoted by placing arrows on the links, as in
figure \ref{figlattice}.

\[
\begin{array}{cc}
\refstepcounter{figure}
\label{figlattice}
\epsfxsize=.4\linewidth
\epsfbox{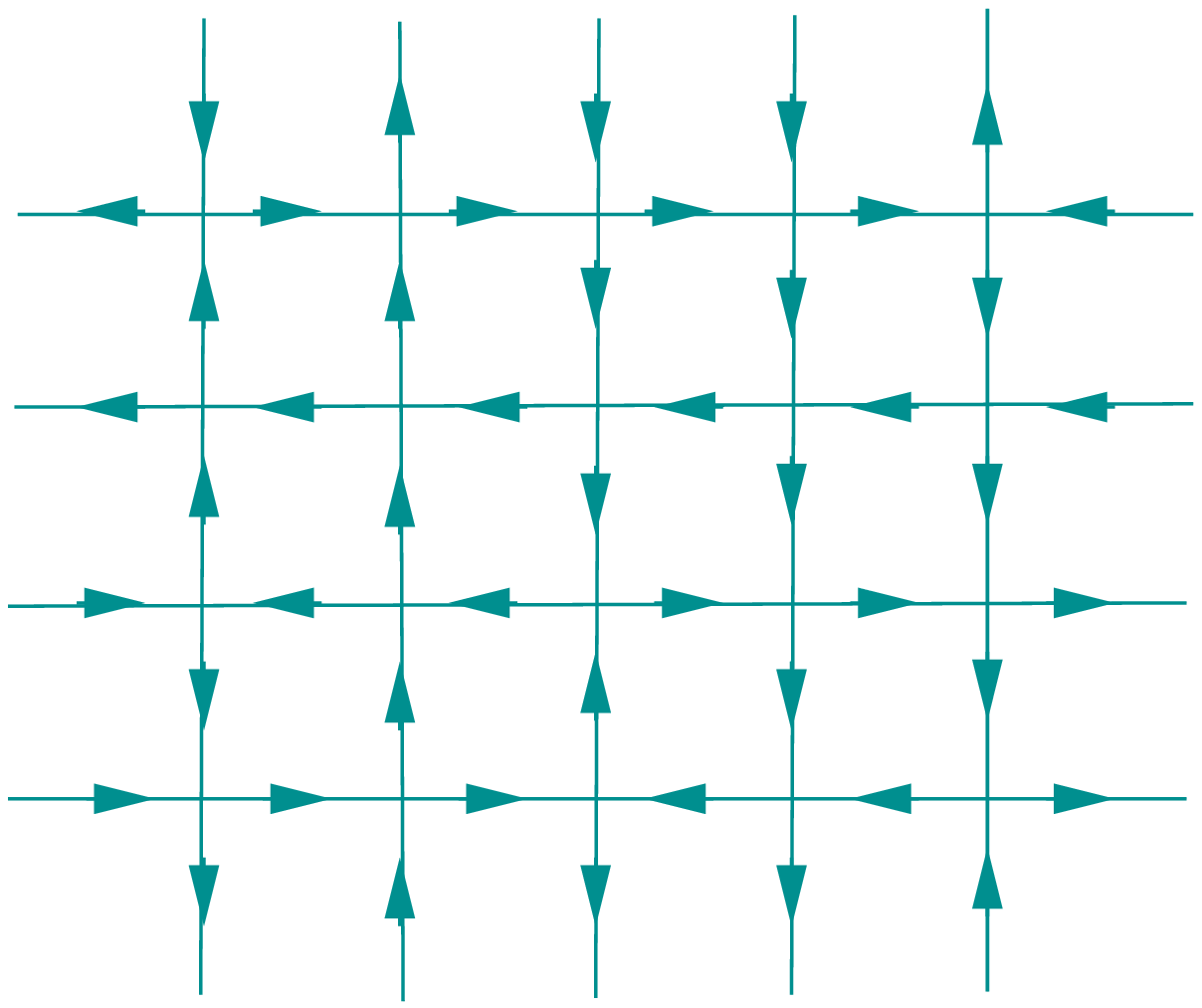}
&
\refstepcounter{figure}
\label{figroots}
\epsfxsize=.4\linewidth
\epsfbox{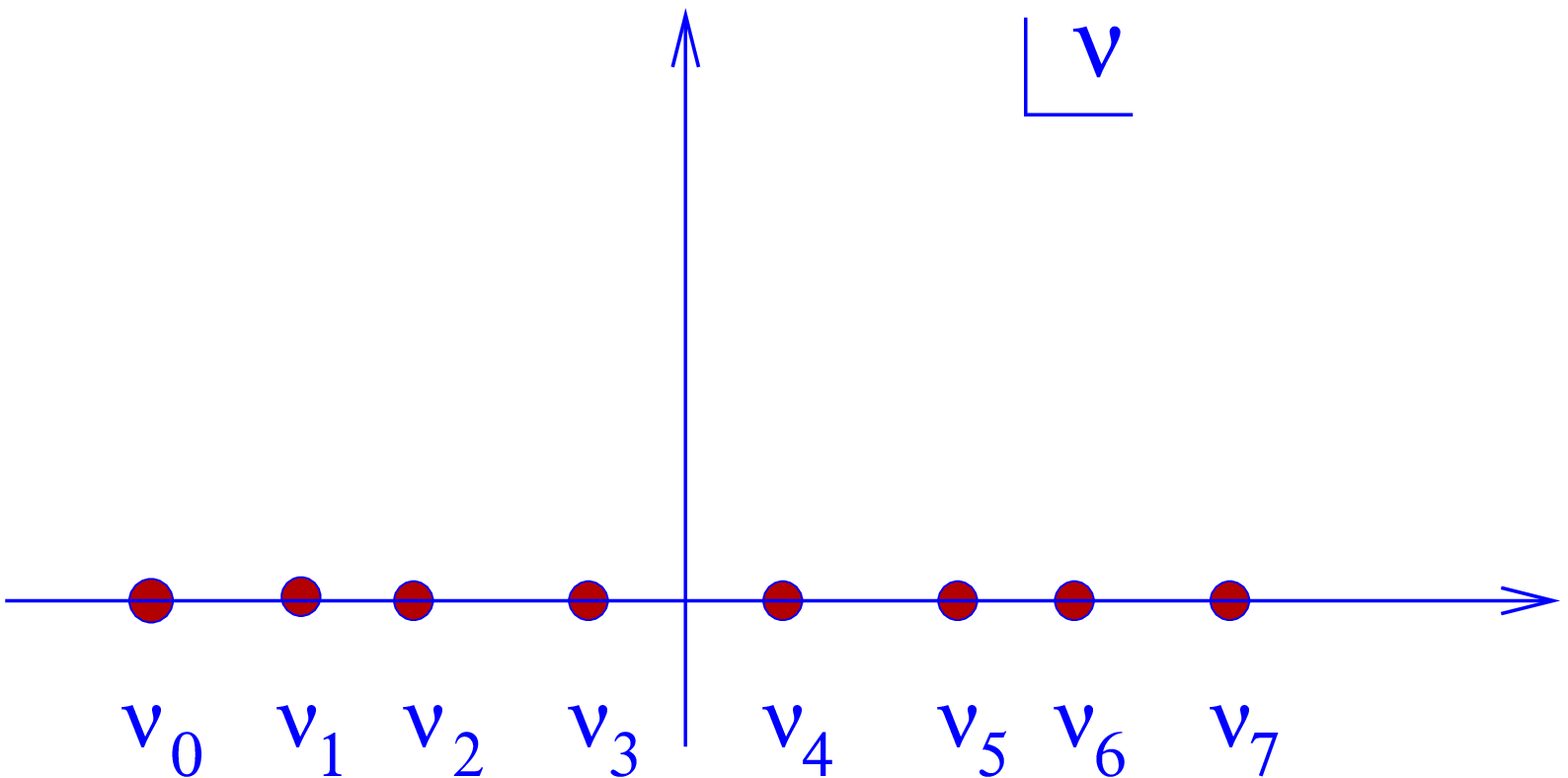}
\\[3pt]
\parbox{.45\linewidth}{\small \raggedright
Figure \ref{figlattice}:  A configuration in the six-vertex
model.}~~&~~
\parbox{.45\linewidth}
{\small \raggedright Figure \ref{figroots}:  The  $\nu_i$'s are real for the
ground state. }
\end{array}
\]
\vskip 0.5cm

\noindent We shall  constrain  the set of configurations as
follows:
\begin{itemize}
\item{Only those configurations of spins which preserve the `flux' of
arrows through each vertex are
allowed;}
\item{We  only consider the `zero field'
 six-vertex model, which has  an additional  `4-spin reversal'
symmetry implying the model is invariant under
simultaneous reversal of all arrows.}
\end{itemize}
Thus the `local Boltzmann weights', numbers assigned to each vertex
depending on the spins next to that vertex, can be parameterised
in terms of just three quantities:
{\small
\[
W\!\left[\matrix{&\uparrow&\cr
                 \rightarrow&&\rightarrow\cr
                 &\uparrow&}\right]=
W\!\left[\matrix{&\downarrow&\cr
                 \leftarrow&&\leftarrow\cr
                 &\downarrow&}\right]=a~,
\]
\[
W\!\left[\matrix{&\downarrow&\cr
                 \rightarrow&&\rightarrow\cr
                 &\downarrow&}\right]=
W\!\left[\matrix{&\uparrow&\cr
                 \leftarrow&&\leftarrow\cr
                 &\uparrow&}\right]=b~,
\]
\[
W\!\left[\matrix{&\uparrow&\cr
                 \rightarrow&&\leftarrow\cr
                 &\downarrow&}\right]=
W\!\left[\matrix{&\downarrow&\cr
                 \leftarrow&&\rightarrow\cr
                 &\uparrow&}\right]=c~.
\]
}

\noindent
In fact the overall normalisation factors out trivially from all
calculations, and the remaining two degrees of
freedom can be  parametrised  using the two variables
\eq
   \nu:~~\mbox{the spectral parameter}~,~~~~ \eta:~~\mbox{the anisotropy}
\en
as
\eq
a=\sinh(i \eta- \nu)~,~~b=\sinh(i\eta+\nu)~,~~c=\sinh(2i\eta)~.
\en
The relative probability of finding any given configuration is
simply given by the product of the Boltzmann weights at each vertex.
The six-vertex model introduced  above is a classic
example of an integrable lattice model.

The first quantity of interest - the partition function $Z$ - is the
sum of these numbers over all possible configurations:
\eq
Z=\sum_{ \rule[-.1cm]{0cm}{.285cm}{\rm arrows~}}\prod_{ {\rm sites}}
W\!\!\left[\mbox{\,$\cdot$~\raisebox{1.5ex}{\,$\cdot$~}
\raisebox{-1.5ex}{\hspace{-14.2pt}~$\cdot$~}~$\cdot$\,} \right]\,.
\en
One popular
technique to calculate $Z$ makes use of the so-called transfer
matrix ${\bf T}$, which performs the  sum over one set of horizontal links:
\[
{\bf T}^{\{ \alpha' \}}_{ \{\alpha \}}\,(\nu) {}~{}={}~{}
\sum_{\{\beta_i\}}
W\!\!\left[\mbox{$\beta_1$\raisebox{1.5ex}{$\,\alpha'_1$}
\raisebox{-1.5ex}{\hspace{-18pt}
$\alpha_1$}$\,\beta_2$} \right] W
\!\!\left[\mbox{$\beta_2$\raisebox{1.5ex}{$\,\alpha'_2$}
\raisebox{-1.5ex}{\hspace{-18pt}
$\alpha_2$}$\,\beta_3$} \right]
\dots\,
W
\!\!\left[\mbox{$\beta_N$\raisebox{1.5ex}{$\,\alpha'_N$}
\raisebox{-1.5ex}{\hspace{-22pt}
$\alpha_N$}$\,\beta_1$} \right]~.
\]
In terms of ${\bf T}$ the  partition function is given by
\eq
Z=\mbox{Trace}\left[{\bf T}^M\right]~.
\en
In the limit  $M\to \infty$ with $N$ finite
the  free energy per site
 can then be obtained as
\eq
f=\fract{1}{N \! M}\ln Z=
  \fract{1}{N\! M}\ln \mbox{Trace}\left[{\bf T}^M\right]\sim\fract{1}{N}
\ln t_0~,
\en
where  $t_{0}$ is the ground-state eigenvalue of ${\bf T}$.
The problem is thus
reduced to the  determination  of $t_{0}$\,; and since the
model is integrable, there are many
methods to achieve this end.
The detailed description of these methods goes beyond the scope
of this review, but it is worth mentioning that
they  usually lead  to a set of non-linear constraints
(functional relations or non-linear integral
equations) on $t_0(\nu)$, or, more simply, constraints on  the
zeroes of a related  function  $q_0(\nu)$,
which we shall introduce shortly. These constraints are very
powerful even at finite $N$ but they lead
to a closed expression for the  free energy only in the
thermodynamic limit $M,N \rightarrow \infty$.

Here we
shall sketch  the logical flow of a method
developed by Baxter  (for more details, see \cite{baxterbook}) leading to a
relation  known as the
TQ-system that can also be easily deduced
starting from the ODE side.  This should provide
a first clear hint of what we mean by an ODE/IM
correspondence.

Baxter began by showing that there exists an auxiliary function
$q_0(\nu)$,
\eq
q_0(\nu)=\prod_{i=0}^{n-1}\sinh(\nu-\nu_i)~,
\label{qq}
\en
such that the following `TQ relation' holds
\eq
t_0(\nu)q_0(\nu)=a(\nu,\eta)^N
q_0(\nu+2i\eta)+b(\nu,\eta)^N q_0(\nu-2i\eta)~.
\label{Tq1}
\en
This is perhaps a puzzling step to take: we want to find
$t_0(\nu)$ and we now claim that the
relation (\ref{Tq1}), which just
defines $t_0(\nu)$ in terms of another unknown function
$q_0(\nu)$,  will somehow help us in this task. The
explanation is simple: the constraint (\ref{Tq1}) should be combined with a
knowledge of the analytic properties of both $t_0$
and $q_0$. In particular the fact that
 $t_0(\nu)$ and $q_0(\nu)$ are
entire functions of $\nu$  means that a powerful constraint, the Bethe
ansatz system, is immediate. By definition
\eq
q_0(\nu_i)=0~,
\en
and combining this with the relation   (\ref{Tq1})
and the fact that $t_0(\nu)$ is entire   leads to
the  following set of
 Bethe ansatz equations (BAE):
\eq
-1= {a^N(\nu_i,\eta)
\over b^N(\nu_i,\eta)} {q_0(\nu_i+2i\eta) \over
q_0(\nu_i-2i\eta)} \quad ,\quad i=0,1,\dots, n{-}1~.
\label{ba}
\en
This already looks to be a more serious constraint than (\ref{Tq1}),
but this is not yet sufficient
since (\ref{ba}) has many sets of solutions $\{\nu_i\}$.
However, one can argue that
there is a particular solution, with $n=N/2$ and all of the $\nu_i$ real
(depicted in figure~\ref{figroots}), which uniquely corresponds to
the ground-state eigenvalue of ${\bf T}$, i.e.  $t_0$.

To prepare the ground  for the connection
with differential equations, let us define
\eq
\lambda_i=e^{2 \nu_i}~~~,~~\omega=-e^{i 2 \eta}~,
\en
so that the Bethe ansatz equations~(\ref{ba}) become
\eq
-1=
\left({1+\lambda_i/\omega  \over 1+ \lambda_i\omega}\right)^N
\ \prod_{n=0}^{N/2-1}
{ (\lambda_n- \lambda_i \omega^2)  \over (\lambda_n- \lambda_i
\omega^{-2}) }~.
\en
We should mention one final detail:
currently correspondences with ordinary differential equations have
only been established with integrable models in
their continuum, conformal limits. For the six-vertex model
this limit is found by  sending
\eq
N \rightarrow \infty~~\mbox{ and} ~~a=e^{\pi  \nu/2 \eta}  \rightarrow 0~,
\en
with   $a N$ kept finite. Sending
\eq
a \rightarrow  0 ~~~\mbox{  as}~~~ a \rightarrow a/N
\en
one discovers that the $\lambda_i $'s with
$i \ll \ln N $ rescale to zero as
\eq
 \lambda_i \sim E_i  a^{4 \eta/ \pi} \sim
E_i N^{-4 \eta/ \pi}\,.
\en
For $\pi/4 <\eta<\pi/2$ the limiting product converges with no need
for extra regulating factors, and one obtains
\eq
-1=
\prod_{n=0}^{\infty}
{ (E_n- E_i \omega^2)   \over (E_n- E_i \omega^{-2}) }~.
\label{baco}
\en
Finally, the  six-vertex model can be generalised to incorporate
twisted boundary conditions without losing integrability.
The twist is introducing by making the following replacement in the
definition of the original model:
\eq
W
\!\!\left[\mbox{$\beta_N$\raisebox{1.5ex}{$\alpha'_N$}
\raisebox{-1.5ex}{\hspace{-22pt}
$\alpha_N$}$\,\beta_1$} \right]
\rightarrow
e^{i (\delta_{\beta_1, \uparrow} - \delta_{\beta_1,\downarrow})\phi}\,W
\!\!\left[\mbox{$\beta_N$\raisebox{1.5ex}{$\alpha'_N$}
\raisebox{-1.5ex}{\hspace{-22pt}
$\alpha_N$}$\,\beta_1$} \right]~.
\en
In  the conformal limit  described above the introduction of the twist
 leads to the more general BAE
\eq
-1= e^{i 2 \phi}
\prod_{n=0}^{\infty}
{ (E_n- E_i \omega^2)   \over (E_n- E_i \omega^{-2}) }~.
\label{baegen}
\en
We shall shortly see how to derive  the same set of equations
 starting from a differential equation.

\resection{$\PT$-symmetric quantum mechanics }
Before we go into the details of the
precise link between ODEs and integrable models
we shall first give a description of the non-Hermitian quantum
mechanical problems mentioned in the prelude.

Researchers working on integrable 1+1 dimensional massive
quantum field theories have become quite accustomed to
 systems described by   non-Hermitian
Hamiltonians which nevertheless possess, at least in
a certain range of the parameters, a real and
positive energy spectrum. A standard example,
which has been studied since the mid-1990s, is the
scaling Lee-Yang model (see, for example,~\cite{Cardy:1985yy}).
This is the theory of a massive
relativistic bosonic field $\phi$ with cubic
interaction $g\phi^3$. Along with the $\lambda \phi^4$ case, it is
one of the standard text-book examples used for practising Feynman diagram
technology~\cite{Collins}. However, since the potential is
unbounded from below, the Lee-Yang model with
real $g$  suffers from obvious pathologies. Surprisingly,
at  least in the  1+1 dimensional integrable
case, these pathologies disappear in the fully
renormalised scaling limit at   $g=i |g|$. We invite the
interested reader to consult~\cite{Cardy:1985yy,Cardy:1989fw,Yurov:1989yu}
to appreciate various aspects of this phenomenon, and to note
in particular
the striking similarity between the plots in figures~6 and 7
in Ref.~\cite{Yurov:1989yu}
and figure~\ref{fig1}a below, which shows the energy
spectrum of a non-Hermitian quantum-mechanical problem, (\ref{BBE}).

Indeed, it was the study of  the 1+1
dimensional  Lee-Yang model that led Bessis and Zinn-Justin to the
following, rather simpler, question: is the energy spectrum
associated with the Schr\"odinger equation
\eq
-\frac{d^2}{dx^2}\psi(x)+ix^3\psi(x)=E_n\psi(x)~,~~~
\psi(x) \in L^2(\RR)
\label{ly}
\en
real?
Perturbative and numerical studies led
Bessis and Zinn-Justin to conjecture that the spectrum $\{E_n \}$
is indeed real, and positive~\cite{BZJ}.
But how does one  prove this statement analytically?

Later, in 1997, Bender and Boettcher~\cite{BB} considered
 the spectrum of the
 following generalisation of Bessis and Zinn-Justin's problem:
\eq
-\frac{d^2}{dx^2}\psi(x)-(ix)^N\psi(x)=E_n\psi(x)~,~~
\qquad\mbox{($N$ real, $>0$)}
\label{BBE}
\en
with the  associated boundary conditions to be specified
shortly. Again, extensive numerical checks led the authors of~\cite{BB}
to  a precise  reality conjecture. They also remarked  that both equations
(\ref{ly}) and (\ref{BBE}) are invariant under
the action of the operator $\cal PT$:
\bea
{\cal P} & \mbox{(parity reflection):}~~&x \rightarrow -x~,~p \rightarrow -p~,
\nn \\
{\cal T} & \mbox{(time reversal):}~~~~~& x \rightarrow ~x~,~p \rightarrow -p~,~ i \rightarrow -i~.
\nn
\eea
The special class of non-hermitian quantum systems with $\cal PT$ symmetry
has only recently started to be studied in depth.  These theories
are mathematically interesting, and they may have an important r{\^o}le in
physics~\cite{Bender:1999ek}.

Returning to the ODE~(\ref{BBE}), we note  that there appear to be a
couple of problems with the
generalisation. Firstly, for non-integer values of $N$ the
`potential' $ -(ix)^N $ is not single-valued
and we need to add a branch cut, which we shall put
 along the positive imaginary $x$-axis:
\[
\begin{array}{c}
\refstepcounter{figure}
\label{figray}
\epsfysize=.24\linewidth
\epsfbox{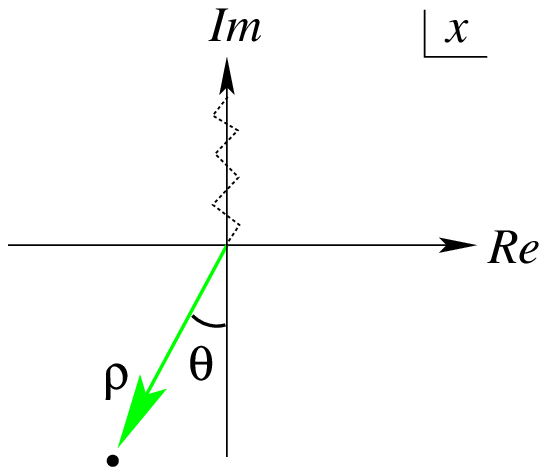}
\\[3pt]
\parbox{.7\linewidth}{\small ~~~Figure~\ref{figray}: The
branch cut, and a ray in the
complex $x$-plane.
}
\end{array}
\]
Secondly, when $N$  reaches $4$  the `potential'
is  $-x^4$, and the eigenvalue problem runs into difficulties
since all the  solutions to (\ref{BBE})
decay algebraically as $|x| \rightarrow
\infty$ on the real axis.
To overcome this problem it is necessary to enlarge the perspective and
treat $x$ as a genuinely complex variable.
Consider solutions to the ODE in the
complex $x$-plane. For most  values of $\arg x\,$, up to an
overall normalisation there will be a unique
exponentially-decaying solution at large $|x|$ (this solution
is called ``subdominant'') and any other
 linearly independent solution will be
exponentially growing (these solutions are called ``dominant'').
However, whenever
\eq
\arg x\,=
\pm\frac{\pi}{N{+}2}~,~
\pm\frac{3\pi}{N{+}2}~,~
\pm\frac{5\pi}{N{+}2}~,~\dots
\en
all solutions decay  algebraically. The lines along which this
occurs are known as `anti-Stokes
lines' and the wedges between
the  anti-Stokes lines are the `Stokes sectors'.
Since all functions involved are analytic, we can continue the
wavefunction along some other contour in
the complex plane, and as long as it does not cross any anti-Stokes
lines the spectrum will remain the same.  At $N=4$ an anti-Stokes line
coincides with the real axis and the correct analytic continuation of
the original problem is achieved by bending the wavefunction contour down
into the complex plane, as shown in figure~\ref{fig5}.

We conclude from this discussion that each  pair of Stokes sectors
in which the  wavefunction is required to decay at large
$|x|$ defines a different
eigenvalue problem.
In the terminology of the WKB method, these are related to
`lateral' connection problems. Contours which
instead join $x=0$ to $x=\infty$ lead to what are called `radial' (or
`central') connection problems. Figure~\ref{fig6} depicts a sample of
the possible lateral and radial wavefunction contours.
\[
\begin{array}{cc}
\refstepcounter{figure}
\label{fig5}
\epsfxsize=.3\linewidth
\epsfbox{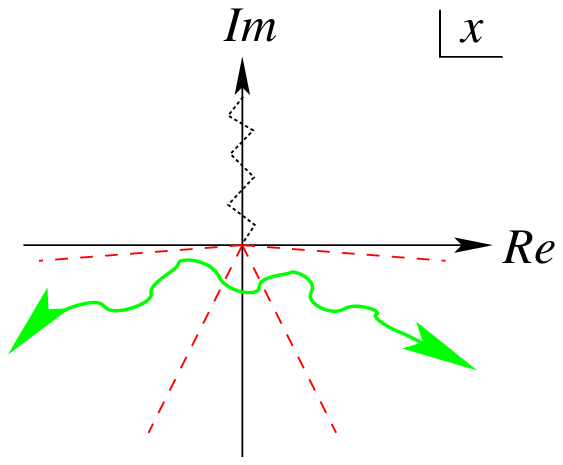}
&
\refstepcounter{figure}
\label{fig6}
\epsfxsize=.3\linewidth
\epsfbox{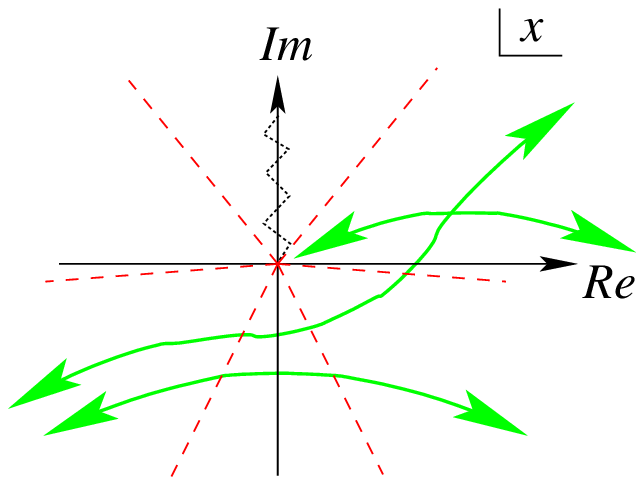}
\\
\parbox{.4\linewidth}{\small \raggedright
Figure \ref{fig5}:  A possible wavefunction contour for $N>4$. } &
\parbox{.4\linewidth}
{\small \raggedright Figure \ref{fig6}:  Some further quantization
contours. }
\end{array}
\]
\vskip 0.35cm
Questions in $\PT$-symmetric quantum mechanics are all
related to lateral problems, with
one particular pair of Stokes sectors selected.
The first eigenproblem to arise in the ODE/IM correspondence
was of radial type~\cite{Dorey:1998pt},
with the contour defined along the positive real axis.
However it turns out that  the $\PT$-symmetric quantum
mechanical problems (\ref{BBE})
are intimately related with certain radial problems,
in a way that will
be discussed in the next-but-one section.
\resection{Numerical evidence}
\label{num}
We now present some numerical evidence concerning $\PT$-symmetric problems of (\ref{BBE}). In
figure~\ref{fig1}a part of the spectrum of~(\ref{BBE}) is plotted. This plot, together with figures
\ref{fig1}b-d, was obtained in~\cite{Dorey:1999uk} via  a  non-linear integral equation. The results
shown in the first plot had previously been obtained in~\cite{BB,Bender:1998gh} by a direct numerical
treatment of the differential equation in the complex plane.

\[\begin{array}{ll}
\epsfxsize=.4\linewidth\epsfysize=.4\linewidth\epsfbox{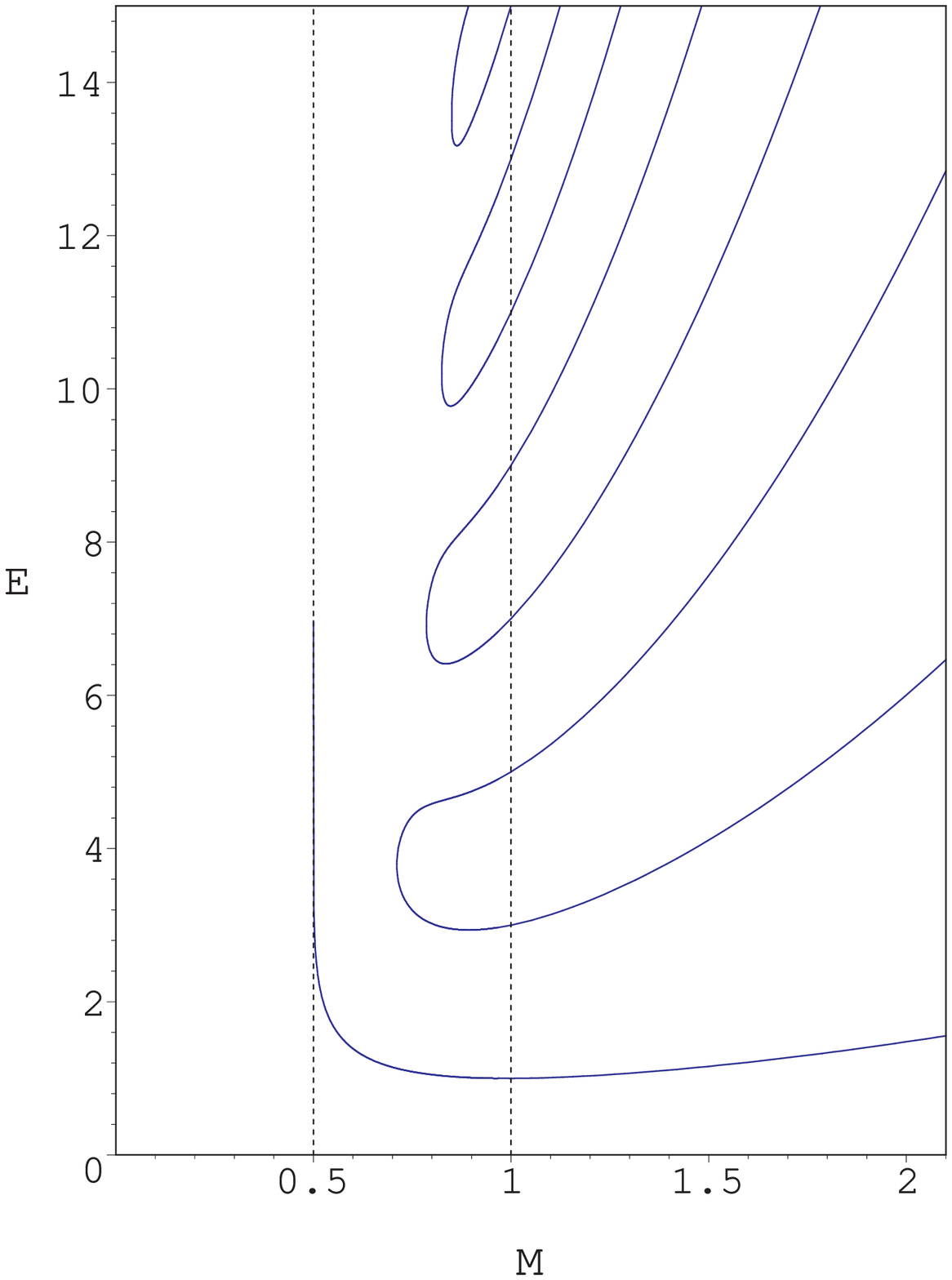}
\refstepcounter{figure}
\label{fig1}
{}~&
\epsfxsize=.4\linewidth\epsfysize=.4\linewidth\epsfbox{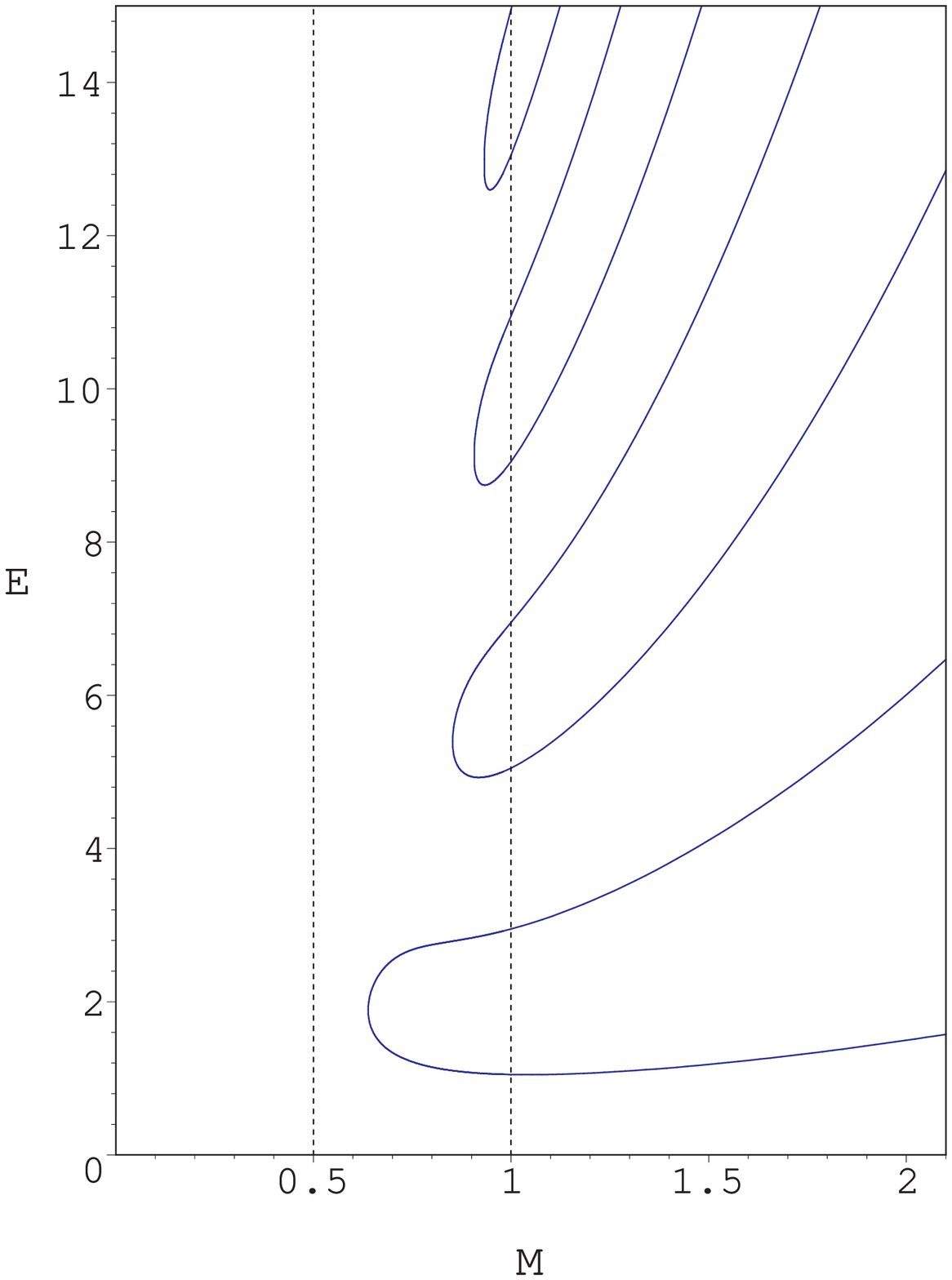}
\\
\parbox[t]{.4\linewidth}{\quad~~~\small Fig.~\ref{fig1}a: $l=0$.}
{}~~&~
\parbox[t]{.4\linewidth}{\quad~~\small  Fig.~\ref{fig1}b: $l=-0.025$.}
\\[11pt]
\epsfxsize=.4\linewidth\epsfysize=.4\linewidth\epsfbox{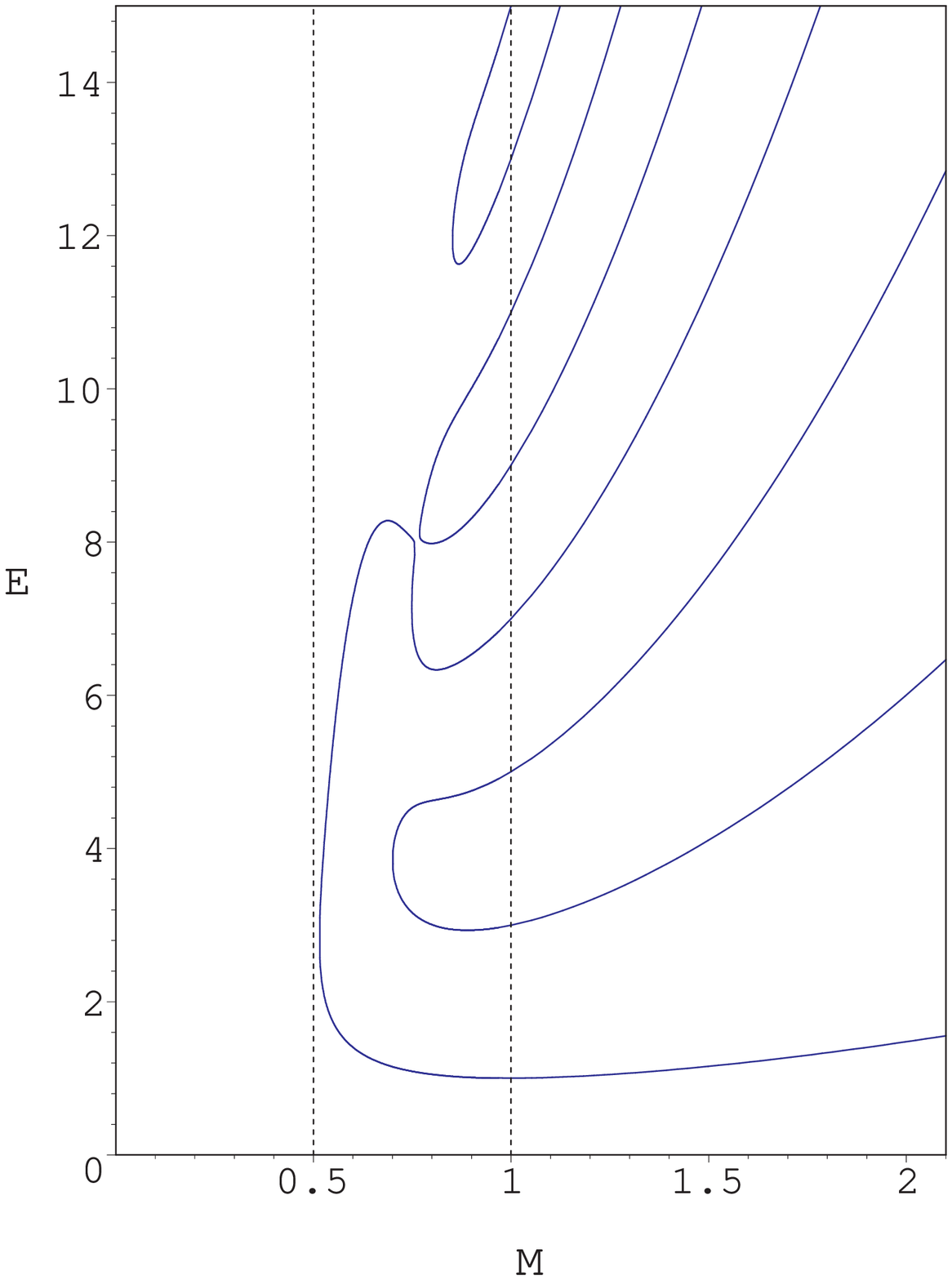}
{}~&
\epsfxsize=.4\linewidth\epsfysize=.4\linewidth\epsfbox{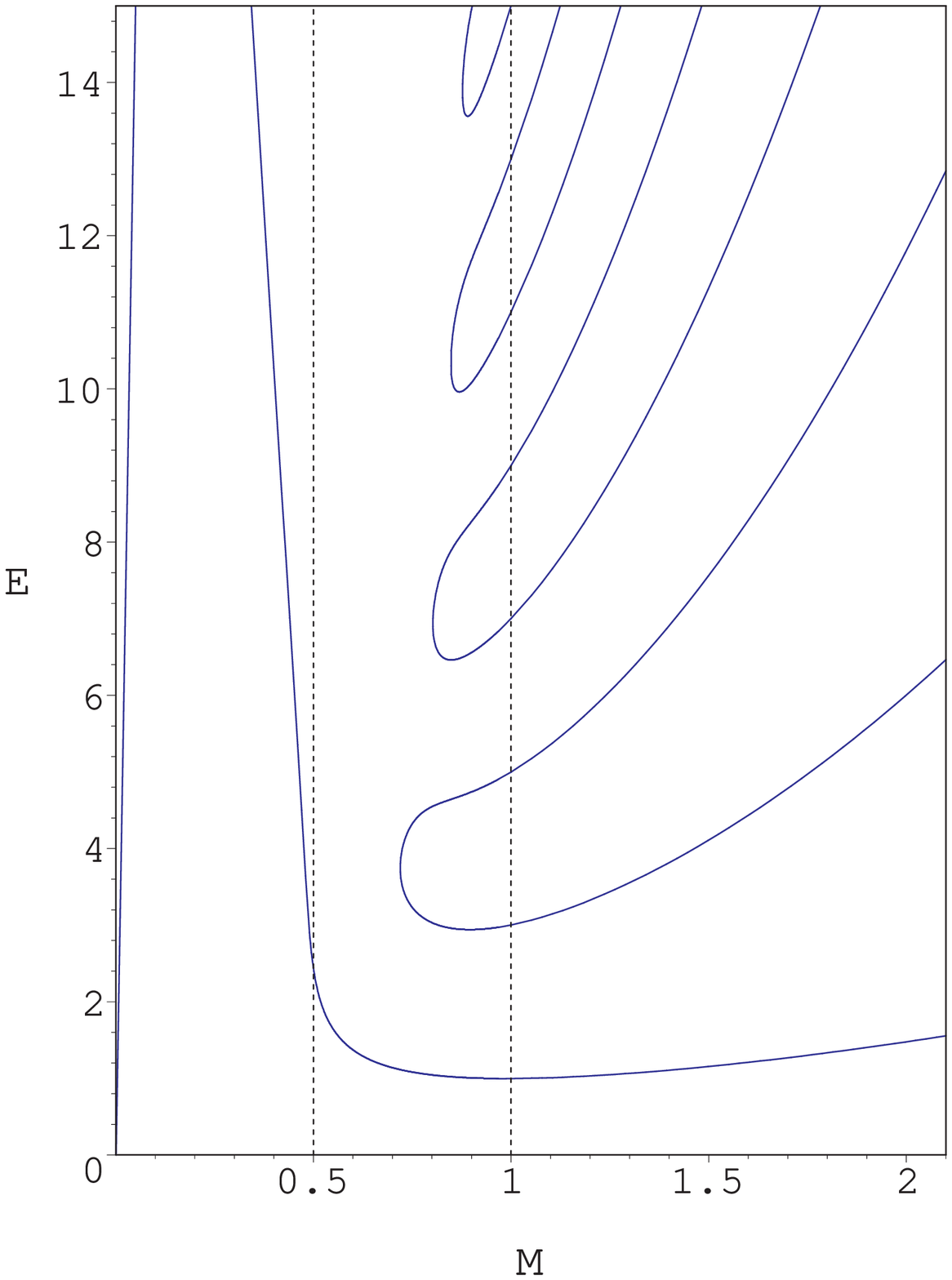}
\\
\parbox[t]{.4\linewidth}{\quad~~~\small Fig.~\ref{fig1}c:
$l=-0.001$.}
{}~~&~
\parbox[t]{.4\linewidth}{\quad~~\small  Fig.~\ref{fig1}d: $l=0.001$.}
\\[8pt]
\multicolumn{2}{l}%
{\qquad\qquad\parbox{.7\linewidth}%
{\small Figure \protect\ref{fig1}: Eigenvalues of the
Hamiltonian
$ p^2-(ix)^{2M}+l(l{+}1)x^{-2}$\,.} }
\end{array}
\]

{}From figure \ref{fig1}a, for $N\equiv 2M \ge 2$ the spectrum is, within
our numerical precision, real and positive, while as $N$
moves below $2$ an infinite number of energy levels pair off
and become complex. The transition to
infinitely-many complex eigenvalues was interpreted in
\cite{Bender:1998gh}
as a spontaneous breaking of ${\cal PT}$ symmetry\,\footnote{
Since $ [{\cal PT},H]=0$ it follow that if an  eigenfunction $\Psi$
of $H$ is also an eigenfunction of
${ \cal PT}$ with eigenvalue $\lambda$ then
${\cal PT}\; H  \Psi= E^* \lambda \Psi~\equiv~
H \;{\cal PT}\Psi= E \lambda \Psi$ and $E=E^*$.}.

We shall tackle the reality question for $N \ge 2$
analytically. For this purpose it
is convenient to enlarge the perspective by including two
extra parameters, $\alpha$ and $l$, a generalisation that
does not add any extra technical difficulties but  gives a wider
phenomenology. It is also convenient to trade the parameter
$N$ for $M\equiv N/2$.  The more general theory
is~\cite{Dorey:1999uk,Dorey:2001uw}
\eq
-\frac{d^2}{dx^2}\psi(x) - \Big( (ix)^{2M} + \alpha (ix)^{M -1} +{l(l+1)
\over x^2} \Big)
\psi(x)=E_n\psi(x)~.
\label{full}
\en
Even with $\alpha=0$,
for $-1 <l<0$ the additional angular-momentum term has
a remarkable effect on the connectivity of the spectrum, as can be seen
in the middle two plots of figure~\ref{fig1}. Moreover, we
shall show in the next sections, as in~\cite{Dorey:2001uw, Dorey:2001hi},
that the spectrum of (\ref{full}) is
\bea
\bullet&\! \mbox{~~{\em real}~~~~ if}&\alpha<M+1+|2l{+}1|\nn
\\[3pt]
\bullet&\!\! \mbox{{\em positive}~ if}&\alpha<M+1-|2l{+}1|~.\nn
\eea
The proof makes use of integrable model technology. While there will
not be space to go into details below, we remark that these ideas can also
be used to study the way that the energy levels merge to become
complex~\cite{Dorey:2004fk}.

\resection{ Analysis of the  Schr\"odinger equation }
As in~\cite{Dorey:1999uk,Dorey:2001uw},
we start the analysis by considering a related  differential equation
\eq
\left (-\frac{d^2}{dx^2}+ x^{2M}+ \alpha x^{M-1}+ {l(l+1) \over x^2} -E
\right)\phi(x)=0
\label{eq5}
\en
\noindent
with   $x$  and  $E$ possibly complex and $M>0$.

We shall need a couple of
important facts. Firstly, the equation  has a solution $y=y(x,\alpha,E,l)$ such that:
\begin{itemize}
\item $y$ is  entire in  $E$
and $x$ (though, due to the
branch point at $x=0$, $x$ must in
general be considered to live on a suitable cover of the punctured complex
plane)\,\footnote{ The entirety of
$y$ was first proved by Sibuya~(see Ref.~\cite{Sha}). His work concerned
only the case  $l=0$, $\alpha=0$,
$2M\in\mathbb{N}$, but the result also holds for the more general
situation of eq.~(\ref{eq5}), so long as
the branching at the origin is taken into account. In this respect
we should   mention that the
$l=0$, $\alpha=0$, $2M\in\RR^+$ case  was explicitly discussed by Tabara in
\cite{Tabara},
while the generalisation to a potential $P(x)/x^2$
with $P(x)$ a polynomial in $x$ was studied by
Mullin~\cite{Mullin}, and more recently in~\cite{DeRa}. It
is also worth noting that with a change of
variable it is possible to map eq.~(\ref{eq5}) with $\alpha \in \RR$, $l \in
\RR$, $2M \in
\mathbb{Q}^+$ onto particular cases of those treated in~\cite{Mullin}.}
\item for $M>1$, $y$ and  $y'= dy/dx$ admit the following asymptotic
representations
\eq
y\sim {  x^{-M/2-\alpha/2} \over \sqrt{2 i} }
\exp(-\fract{1}{M+1}x^{M+1}) \quad , \quad y'\sim -{
x^{M/2-\alpha/2}\over \sqrt{2 i} } \exp(-\fract{1}{M+1}x^{M+1})
\en
for $|x| \rightarrow \infty$ in any closed sector contained in the
sector $|\arg x\,|<\frac{3\pi}{2M+2}$ (though
extra terms appear for $0< M \le 1.$)
\end{itemize}
\nobreak
Secondly, setting  $x= \rho e^{i \theta}$ as  in figure~\ref{figray} with
$\rho$ real, and denoting  the sector
$\left| \theta - \fract{ k \pi}{M+1} \right| < \fract{\pi}{2M+2}
$ by $\CS_k$ as in figure~\ref{figsecs}, we see that, as
$\rho\to\infty$ with $\theta$ fixed,
\nobreak
\begin{itemize}
\item{
in $\CS_0: y  \rightarrow
 0$~~~~{ ($y$ is subdominant in $\CS_0$)}
}
\item{
at  $\theta = \pm \fract{\pi}{2M+2}$~~ { $y$ decays algebraically} }
\item{
in $\CS_{\pm 1}: y \rightarrow
\infty$~
($y$  is dominant in $\CS_{\pm 1}$)}.
\end{itemize}
 \[
\begin{array}{c}
\refstepcounter{figure}
\label{figsecs}
\epsfxsize=.40\linewidth\
\epsfbox{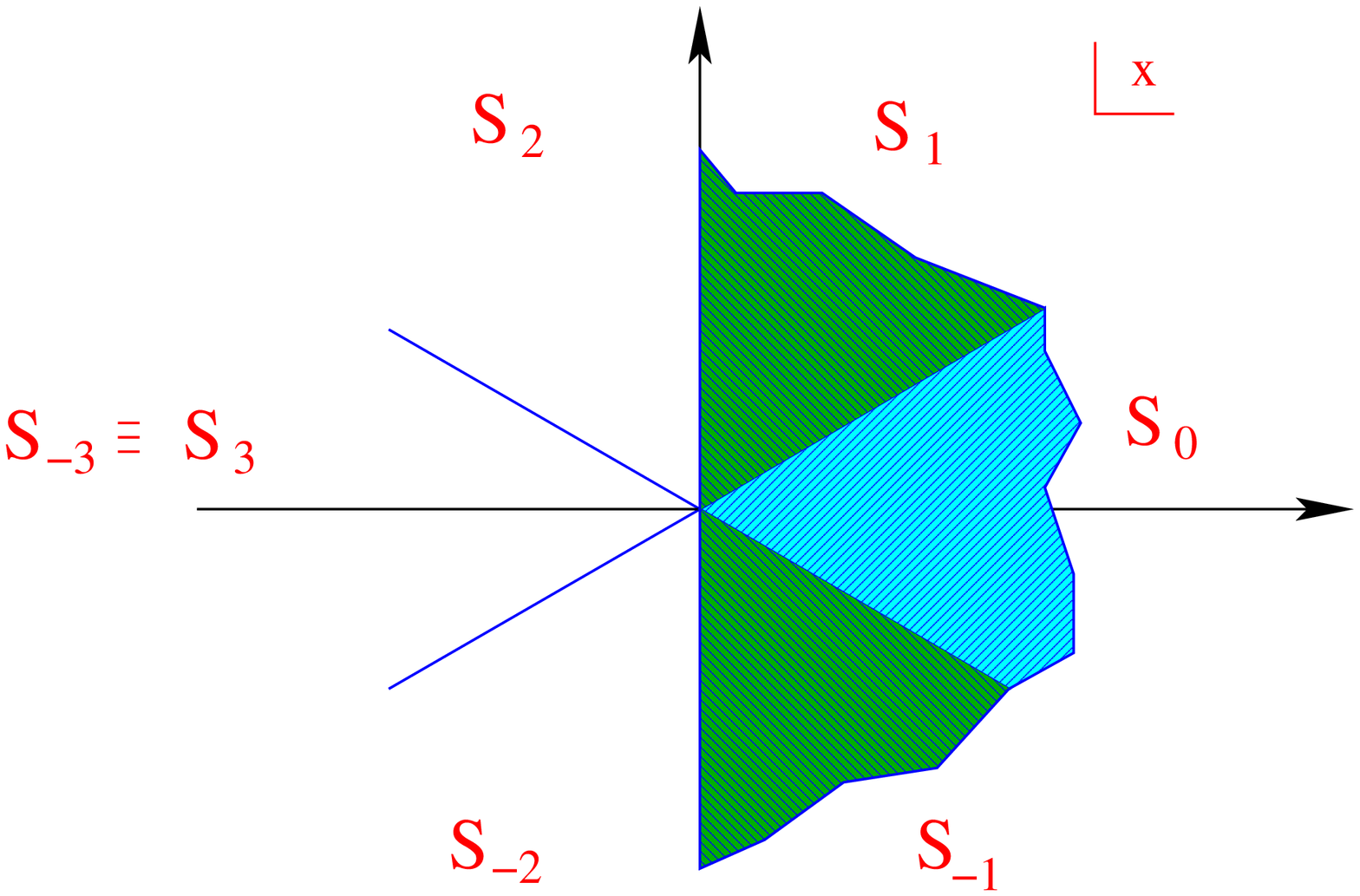}\\
\parbox{.7\linewidth}{\raggedright \qquad Figure~\ref{figsecs}:
    The sectors ${\cal S}_k$ for the potential $x^4$. }
\end{array}
\]

It is then easy to see that the
uniquely-determined solution $y(x,E,\alpha,l)$ defines an  associated set
of functions
\eq
y_k = \omega^{k/2 +  k\alpha/2} y(\omega^{-k}  x,\;\omega^{-2 M k} E,\;
 (-1)^{k}\alpha,\;l) \quad , \quad
\omega= e^\fract{i \pi}{M+1}~,
\en
which are also  solutions of~(\ref{eq5})
for integer $k$, and any pair  $\{y_k,y_{k{+}1} \}$
forms a
basis of solutions.
We can therefore write $y_{-1}$ as a  linear combination of $y_0$  and
$y_{1}$. The result has the form
\eq
T(E,\alpha,l) y_0(x,E,\alpha,l)=y_{-1}(x,E,\alpha,l)+ y_{1}(x,E,\alpha,l)
\,,
\en
where the function $T$ is called a Stokes multiplier.
Keeping $\Re e\,l>-1/2$, the leading behaviour of $y$ near $x=0$ at generic
$E$ is
\eq
y(x,E,\alpha,l) \sim Q(E,\alpha,l ) x^{-l}+\dots~,
\en
and in terms of the shorthand notation
\eq
T^{(\pm)} = T(E, \pm \alpha,l)\, ,~~Q^{(\pm)} =Q^{(\pm)}(E)= Q(E,\pm
\alpha,l)
\en
we find the following relations which intertwine the as-yet
undetermined functions $T^{(\pm)}$ and $Q^{(\pm)}$:
\bea
T^{(+)}(E)Q^{(+)}(E)&=&\omega^{-\fract{2l+1+\alpha}{2}} Q^{(-)}(\omega^{2M} E)+
\omega^{\fract{2l+1+\alpha}{2}} Q^{(-)}(\omega^{-2M}  E)
\label{tqa}
\\[4pt]
T^{(-)}(E)Q^{(-)}(E)&=&\omega^{-\fract{2l+1-\alpha}{2}} Q^{(+)}(\omega^{2M} E)+
\omega^{\fract{2l+1-\alpha}{2}} Q^{(+)}(\omega^{-2M}  E)~. \qquad
\quad
\label{tqb}
\eea
Notice the striking similarity between equations~(\ref{tqa}),
(\ref{tqb}) and the Baxter TQ-system of equation~(\ref{Tq1}).

At the zeroes $\{ E_k(l,\alpha) \}$ of $Q(E)$, the leading behaviour
of $y$ at the origin changes to
\eq
y(x,E_k,\alpha,l) \sim Q(E_k,\alpha,-l{-}1 ) x^{l+1}+\dots~,
\en
and
$y(x,E,\alpha,l)$ decays at the origin as well as at infinity.
This implies that $Q(E)$  is
the spectral determinant encoding the eigenvalues of (\ref{eq5}) for
boundary conditions of radial type.
For $M>1 $, we can use the Hadamard
factorisation theorem to write $Q$ as:
\eq
Q(E,l,\alpha) =Q(0,l,\alpha)
\prod_{n=0}^{\infty} \left( 1- {E \over E_n} \right)~.
\en
Both $ T^{(\pm)}(E) $  and $ Q^{(\pm)}(E) $ are
 entire in $E$, so the LHS of the relevant TQ  equation ((\ref{tqa})
 or (\ref{tqb})) vanishes at
\eq
E = E_k^{(\pm)}= E_k(l, \pm \alpha)~,
\en
and the  following system of equations of Bethe ansatz type for the
 energy spectrum is obtained
\bea
\prod_{n=0}^{\infty} \lf( { E^{(-)}_n - \omega^{-2M} \, E^{(+)}_k  \over
E^{(-)}_n -  \omega^{\;2M} \, E^{(+)}_k}
\ri)
&=& - \omega^{-2l-1 - \alpha } \, , \nn  \\
\prod_{n=0}^{\infty} \lf( { E^{(+)}_n - \omega^{-2M} \,
E^{(-)}_k   \over E^{(+)}_n - \omega^{\;2M} \, E^{(-)}_k }
\ri)
&=& - \omega^{-2l-1 + \alpha }\,~. \label{ba2}
\eea
At  $\alpha=0$, $\{E^{(+)}_k \}  = \{ E^{(-)}_k \}$ and the set of equations (\ref{ba2}) reduces to the
Bethe ansatz system~(\ref{baegen})  of the six-vertex model in its
continuum limit. (For $\alpha\neq 0$, the mapping
is instead to a theory called the three-state
Perk-Schulz model \cite{Suzuki:2000fc,PS}.)

The Stokes multiplier
\eq
T(E,\alpha,l)=W[\,y_{-1}\,, y_1
]\, =
\mbox{Det}
\lf[\matrix{y_{-1}(x)  & y_{1}(x)  \cr
y'_{-1}(x) & y'_{1}(x)}\ri]\,,
\en
given here as a Wronskian,
vanishes if and only if
\eq
W[y_{-1},y_1]=0 ~~~ \leftrightarrow ~~~ y_{-1}~ \mbox{\&}~ y_1 ~~~
\mbox{are linearly  dependent.}
\en
This holds if and only if the ODE has a solution decaying
in the two sectors $\CS_{-1}$  and $\CS_1 $ simultaneously.
Since (\ref{eq5}) is related to the $\PT$-symmetric problem (\ref{full})
by the transformation $x\to x/i$, $E\to -E$, this means that
$T(-E,-\alpha,l)$ is  precisely the spectral determinant
for the generalised
Bender-Boettcher problem (\ref{full}).
Thus the TQ relations (\ref{tqa}), (\ref{tqb}) encode the spectra
of  both  radial and lateral eigenproblems.
\resection{A simple proof of the reality property }
In order to prove the reality and positivity claims made at the end of
section~\ref{num} we return to equation (\ref{tqa})
\eq
T^{(+)} Q^{(+)} =\omega^{-\fract{2l+1+\alpha}{2}} Q^{(-)}(\omega^{2M} E)+
\omega^{\fract{2l+1+\alpha}{2}} Q^{(-)}(\omega^{-2M}  E)\, ,
\en
and  define the zeroes of $ T^{(+)}=T(E,\alpha,l) $  to be $E \in
\{ -\lambda_k\}$.
 Setting $ E=
-\lambda_k $  and using the factorised form for  $ Q^{(-)}$ we get
\eq
\prod_{n=0}^{\infty} \left( { E^{(-)}_n + \omega^{-2M} \, \lambda_k  \over
E^{(-)}_n +  \omega^{2M} \, \lambda_k}
\right)
= - \omega^{-2l-1 -\alpha }\, ,\qquad k=0,1,\dots~.
\en
Since the original $ \PT $   eigenproblem  is invariant under $l \to -1{-}l$  we can assume $ l\ge -1/2 $.
Then each $ E^{(-)}_n $ is an eigenvalue of a Hermitian eigenproblem
associated with $\CH(M,-\alpha,l)$,
and hence is real. It is also easy to show that
the eigenvalues  are  all positive, provided
$\alpha<M{+}2l{+}2$ \cite{Dorey:2001uw}.
Taking the modulus${}^2$, using the reality of the $ E^{(-)}_k, $ and writing
 $ \lambda_k=|\lambda_k|\exp(i\,\delta_k)$,
 we have
\eq
\prod_{n=0}^{\infty} \lf ( { (E^{(-)}_n)^2 + |\lambda_k|^2  + 2  E^{(-)}_n
 |\lambda_k| \cos(\fract{2 \pi}{M+1} + \delta_k)
 \over
(E^{(-)}_n)^2 + |\lambda_k|^2  + 2  E^{(-)}_n
 |\lambda_k| \cos(\fract{2\pi}{M+1} - \delta_k)}
\ri)
= 1\,.
\en
\noindent
For $ \alpha<M+2l+2$, all of the $E^{(-)}_n$ are strictly positive and
each single term in the product is either greater than, smaller than
or equal to one  depending  only on the cosine terms. To match the RHS
we therefore must have
\eq
\cos(\fract{2\pi}{M+1}+\delta_k)=\cos(\fract{2\pi}{M+1}-\delta_k)\,.
\en
Since   $ M > 1 $ the only possibility is
 \eq
\delta_k= n \pi\, ,~~~~n \in \ZZ
\en
and the eigenvalues are indeed real.
Relaxing the condition on $ l $
we have shown the spectrum of~(\ref{full}) is
\bea
\!\bullet&\! \mbox{~~{\em real}~~~~ if}&\alpha<M+1+|2l{+}1|\nn
\eea
and using continuity in $M$ to keep
track of the signs of the eigenvalues (see \cite{Dorey:2001uw}) it can
then be seen that the spectrum is
\bea
\bullet&\!\! \mbox{{\em positive}~ if}&\alpha<M+1-|2l{+}1|~.\nn
\eea
Referring to figure~\ref{fig7}, the spectrum is entirely real for
$(\alpha,l) \in B \cup C \cup D$ and positive for $(\alpha,l) \in D$ .
For $M<1 $, the order of $ Q^{(-)} $  is greater than one, the
Hadamard-factorised
form for $ Q^{(-)}(E) $ no longer has such a
simple form, and the proof breaks down. In fact we know that most of
the $\lambda_k$'s  become complex in this region.

Finally, we remark that while the above constraints on the parameters
$M$, $\alpha$ and $l$ are sufficient they are not necessary,
as can be seen by studying figure~\ref{fig8} for the case of $M=3$.
The full domain of unreality obtained
numerically in \cite{Dorey:2001hi} is shown as the interior
of the curved line, a  proper subset of $A$.
\[
\begin{array}{cc}
\refstepcounter{figure}
\label{fig7}
\epsfxsize=.3\linewidth
\epsfbox{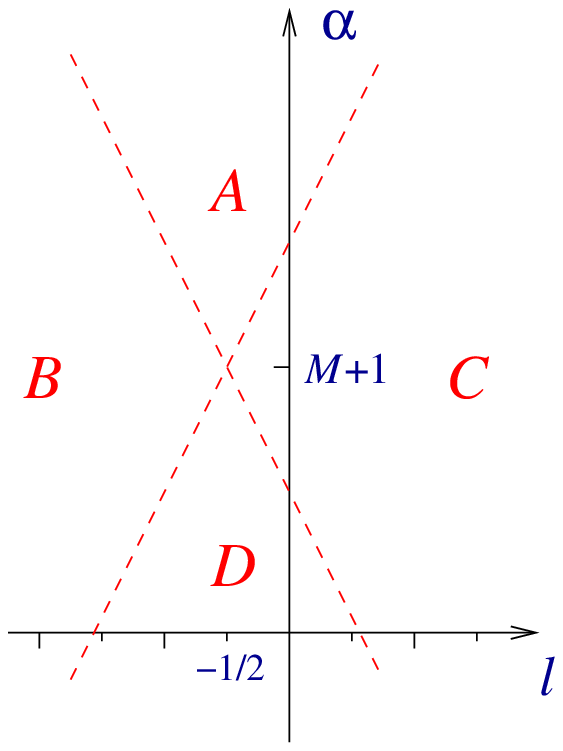}
&
\refstepcounter{figure}
\label{fig8}
\epsfxsize=.3\linewidth
\epsfbox{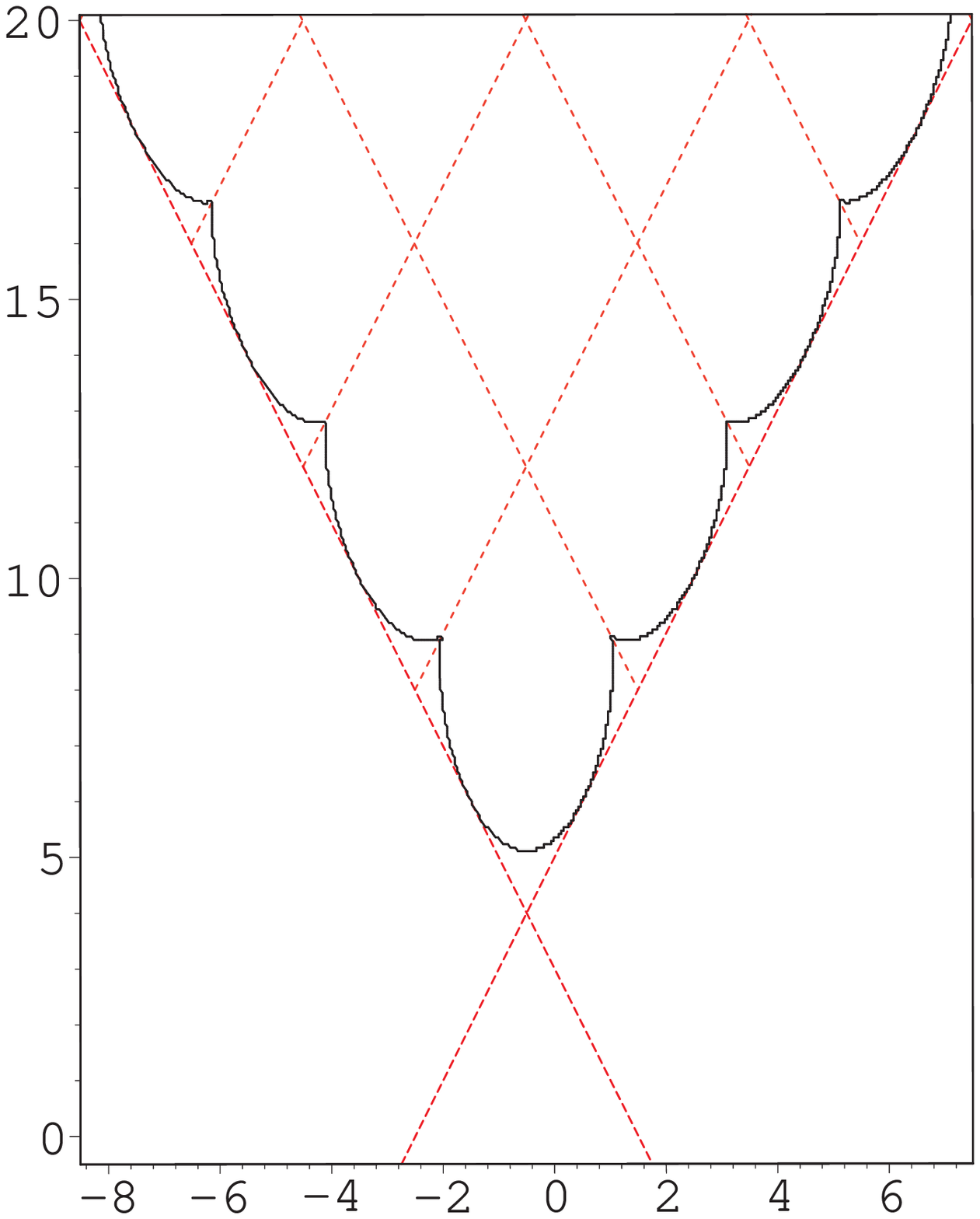}
\\[7pt]
\parbox{.4\linewidth}{\small \raggedright
Figure \ref{fig7}: The `phase diagram' at fixed $M$.  } &
\parbox{.4\linewidth}
{\small \raggedright Figure \ref{fig8}: The domain of unreality
  {\it A} for $ M=3 $ obtained numerically.  }
\end{array}
\]
\vskip 0.5cm
\resection{Conclusions }
We hope to have shown that there is a very interesting relationship
between  the conformal limit of two dimensional integrable models
and the spectral theory of ordinary differential equations.
 The instance described in this review involves the six-vertex model in its conformal limit and an interesting
class  of $\cal PT$-symmetric
 quantum
mechanical systems with complex potential. As an application of this correspondence we have  briefly
sketched the  proof of a conjecture due to Bessis, Zinn-Justin, Bender and Boettcher concerning the
reality of the spectrum of  a particular class of $\cal PT$-symmetric operators.
\section*{Acknowledgments}
RT thanks the organisers for the invitation to speak at the conference.
 We would also like to thank
 Carl Bender, Barry McCoy, Eric Delabaere,
 Junji Suzuki, Yoshitsugu Takei, Andr\'e Voros
  and Miloslav Znojil for useful conversations. This work
was partly supported by the EC network ``EUCLID", contract number HPRN-CT-2002-00325, and partly by a NATO
grant PST.CLG.980424. PED was also supported  by JSPS/Royal Society grant and RT by RIMS.
 CD thanks the Australian Research Council for financial support.
%
%

\end{document}